\documentclass[preprint,showpacs,preprintnumbers,amsmath,amssymb]{revtex4}

\usepackage{graphicx,subfigure}
\usepackage{dcolumn}
\usepackage{bm}

\newcommand{\be}{\begin{equation}}
\newcommand{\ee}{\end{equation}}
\newcommand{\bea}{\begin{eqnarray}}
\newcommand{\eea}{\end{eqnarray}}
\newcommand{\mev}{\rm MeV}

\begin{document}

\bibliographystyle{apsrev}

\preprint{UAB-FT-537}

\title{Primordial Nucleosynthesis as a Test of the Friedmann Equation in the Early Universe} 

\author{Eduard Mass{\'o}} 
\email[]{masso@ifae.es}
\author{Francesc Rota}
\email[]{rota@ifae.es}

\affiliation{Grup de F{\'\i}sica Te{\`o}rica and Institut 
de F{\'\i}sica d'Altes
Energies\\Universitat Aut{\`o}noma de Barcelona\\ 
08193 Bellaterra, Barcelona, Spain}


\date{\today}

\begin{abstract}
In the standard hot big bang model, the expansion of the early universe is given by the Friedmann equation with an energy density dominated by relativistic particles. Since in a variety of models this equation is altered, we introduce modifications in the Friedmann equation and show that we can constrain them using big bang nucleosynthesis data. When there is no neutrino/antineutrino asymmetry these modifications are tightly bounded but in presence of an asymmetry the bounds become much looser. As an illustration, we apply our results to a model where the second and third families couple to gravity differently than the first family (non-universal gravity).

\end{abstract}

\maketitle


\section{Introduction and Conclusions}
\label{intro}
The framework of our present understanding of the evolution of the universe is based on the hot big bang model. Assuming the properties of homogeneity and isotropy of the universe, the space-time metric has to be the Robertson-Walker metric
\be
ds^2 = dt^2 - a^2(t) \left\{ \frac{dr^2}{1-k\,r^2}  + r^2\,d\theta^2 + r^2\,\sin^2\theta\,d\phi^2 \right\}
\label{metric}
\ee
Here, $a(t)$ is the scale factor, and $k=+1,0$ or $-1$ corresponds to a closed, flat or open universe. The other fundamental ingredient for the dynamics of the universe are the Einstein equations. These equations, along with the metric (\ref{metric}), and in a universe containing a perfect fluid of total energy density $\rho$ and pressure $p$, imply the Friedmann equation (FE),
\be
H^2 = \frac{8\pi\, G}{3}\,\rho - \frac{k}{a^2}
\label{friedmann_1}
\ee
with $H\equiv\dot{a}/a$ the Hubble expansion rate. One also deduces the energy-momentum conservation equation
\be
\dot{\rho} = -3\,(\rho + p)\, H
\label{conservation}
\ee

To have a predictive model we need, in addition to Eqs. (\ref{friedmann_1}) and (\ref{conservation}), to use other inputs, like the density of baryons, the density of dark matter and energy, etc. However, Eqs. (\ref{friedmann_1}) and (\ref{conservation}) are very basic equations directly determining the evolution of the universe. In this sense, it is worth to test them as much as we can, since they could be altered by the presence of new physics \cite{Binetruy:1999ut,Cline:1999ts,Rafelski:2002gq,Giudice:2000ex,Freese:2002sq}. In principle, we do not expect a radical change from the form (\ref{friedmann_1}), but the possibility of some alteration is open. 

In this article we will concentrate on the FE (\ref{friedmann_1}), specifically on whether the evolution of the universe at its earliest times is given exactly by the FE. We will constrain potential modifications to the FE in the period of big bang nucleosynthesis (BBN), since this is the earliest epoch where we have observational information. We will show that we can test this equation with the present available data. 

The period of BBN starts at $T \sim 1$  MeV with the decoupling of the weak interaction processes that allow $n \longleftrightarrow p$ inter-conversion and ends with the nuclear fusion production of the light elements, at about $T \sim 50$ keV.

It is an easy exercise to particularize the FE (\ref{friedmann_1}) in the radiation era, where BBN happens. For $T\ll m_\mu$, the relevant energy density appearing in (\ref{friedmann_1}) is given by
\be
\rho = \rho_o \equiv \rho_\gamma + \rho_{\nu_e} + \rho_{\nu_\mu} + \rho_{\nu_\tau} + \rho_e
\label{rho_t}
\ee
In (\ref{rho_t}) we have the contribution from photons, 
\be
\rho_\gamma = g_\gamma\,\int\frac{d^3k}{(2\pi)^3}\,\frac{E}{\exp(E/T) - 1} = 2\,\frac{\pi^2}{30}\,T^4 
\label{rho_g}
 \ee
with the degrees of freedom $g_\gamma = 2$ and $E = |\vec k|$. We also have the contribution of the three neutrino species (neutrinos plus antineutrinos) and of electrons plus positrons
\be
\rho_i = g_i\,\int\frac{d^3k}{(2\pi)^3}\,\frac{E}{\exp[(E-\mu_i)/T] + 1} + 
g_i\,\int\frac{d^3k}{(2\pi)^3}\,\frac{E}{\exp[(E+\mu_i)/T] + 1}
\label{rho_i}
\ee
with the degrees of freedom $g_\nu = 1$ and $g_e = 2$, $E = \sqrt{k^2+m_i^2}$, and we allow for a particle-antiparticle asymmetry by introducing a chemical potential $\mu_i$. 

We write the FE in the radiation dominated epoch as
\be
H^2 = \frac{8\pi G}{3}\,\rho_o
\label{friedmann_2}
\ee
with $\rho_o$ given in (\ref{rho_t}) and where we have neglected the $k$-curvature term in (\ref{friedmann_1}).

The article is organized as follows. First, in section \ref{testing}, we add a new contribution to $\rho_o$ and constrain it when the neutrino/antineutrino asymmetry vanishes, $\mu_\nu = 0$. We show that the new contribution is tightly constrained. In particular we follow the suggestion of the model presented in \cite{Rafelski:2002gq}, and introduce modifications in $G$. More specifically, we let the second and third generation neutrinos to enter in Eq. (\ref{friedmann_2}) with a coupling $G_2$ and $G_3$, respectively, different from $G$. As a consequence of our general study, we can place strong constraints on this model. In section \ref{degenerate} we allow for $\mu_\nu \neq 0$ and show that one may have larger departures from Eq. (\ref{friedmann_2}). However, even in this general case, we will be able to rule out the model of \cite{Rafelski:2002gq}.

\section{Testing the Friedmann equation with nondegenerate BBN}
\label{testing}
Modifications to the FE arise in a variety of contexts. For example, in brane-world models with extra dimensions one obviously finds departures from standard general relativity, and hence the FE is altered. One may have radical changes in this equation when compared to the standard equation (\ref{friedmann_1}) \cite{Binetruy:1999ut} but then it is difficult to reconcile the theoretical predictions with observations. Thus, the hard work is in the struggle to find scenarios where one gets an evolution equation that at first approximation is close to the FE, with modifications that can be tuned to be small \cite{Cline:1999ts}. How small they must be ? We will give an answer using certain type of alterations of the FE (\ref{friedmann_2}), and constraining them in the radiation period using BBN. In this section we set $\mu_{\nu_i} = 0$, but in the next section we will relax this condition.

Apart from brane-world models, we have also found inspiration in two other recent papers, where the general FE (\ref{friedmann_1}) is assumed to be valid, but where there are contributions to $\rho$ not having exactly the form (\ref{rho_g}), valid for radiation. First, in \cite{Giudice:2000ex}, the authors show that the reheating temperature $T_r$ can be as low as  $T_r\sim 0.7$ MeV (when BBN starts). For $T$ above $T_r$ we would have the decay of coherent oscillations of the inflaton, or the decay of a modulus or similar field. If $T_r$ indeed happens to be as low as speculated in \cite{Giudice:2000ex}, it is conceivable that in the period of BBN, for $T<T_r$, apart from having radiation dominating the energy density $\rho$, we might still have some traces of the period $T>T_r$ amounting to some small contribution to $\rho$ that should be added to the r.h.s. of Eq. (\ref{friedmann_2}). Second, in another context, the so-called Cardassian expansion model has been shown to lead to a model in which the universe is flat, matter dominated and accelerating \cite{Freese:2002sq}. This model simply adds a new contribution to the energy density on the r.h.s. of the FE (\ref{friedmann_1}), which would dominate at a late epoch in the evolution of the universe. In our article we consider similar modifications but, as we said, we constrain them in the BBN period.

With all this in mind, we add a new term to Eq. (\ref{friedmann_2}) 
\be
H^2 = \frac{8\pi}{3}\,G \left[\rho_o + \lambda\,\left(\frac{T}{0.1\,\mev}\right)^\gamma \right]
\label{fried_mod_1}
\ee
with $\gamma$ and $\lambda$ two arbitrary parameters, that we shall bound. Even if the type of modification that we introduce in (\ref{fried_mod_1}) is arbitrary, some values of the parameter $\gamma$ have a particular physical significance. For example, the contribution of a term with $\gamma = 0$ would correspond to a cosmological constant. For $\gamma = 4$, the new term would be of the form of extra relativistic particles, and for $\gamma = 3$ it would correspond to extra massive particles.

The new term in the FE modifies the expansion rate, which plays a crucial role in the dynamics during BBN. The universe, just before this period, consists of a hot dense plasma made of $n$, $p$, $e^{\pm}$, $\nu$ and $\gamma$. Neutrons and protons are in equilibrium through weak interactions. This equilibrium is lost when the universe is expanding so fast that particles cannot interact with each other. It happens when the expansion rate $H$ equals the weak interaction rate $\Gamma_W$, a process that is known as the freeze out of the weak interaction. In general, any modification of the expansion rate changes the time of the freeze out.

After freeze out, the temperature of the cooling universe will eventually be low enough to permit the formation of the lightest nuclides. In order to calculate the abundances of these nuclides is important to know how many neutrons and protons are there when the nuclear reactions begin. The value $n/p$ has its equilibrium value, which decreases with time $(m_n > m_p)$, until the freeze out of weak interactions. Since then, neutrons are not destroyed or created by two body reactions, but can still be destroyed by neutron decay. Anyway, to know how many neutrons are there, one has to know at which temperature the weak reactions freeze out, because depending on the freeze out temperature, there will be more or less neutrons available to form the lightest elements.

Before we do the numerical analysis, we can understand some qualitative features of the effect of the new term in (\ref{fried_mod_1}). We can see that for positive values of $\lambda$, its effect is to increase the expansion rate of the universe. Then, as we have said, the weak interaction will freeze out before the time of the freeze out in the standard case, and there will be more neutrons present in the universe at the moment the nuclides begin to form. So, the presence of a term with $\lambda > 0$ will increase the light element yields .

The effect of a $\lambda < 0$ term would be the contrary, but in that case there is the problem that for certain values of $\lambda$ and $\gamma$, $H^2$ becomes negative, which is physically unacceptable. This is indeed what happens. If we look at (\ref{fried_mod_1}), we can see that for large values of $\gamma$ ($\gamma > 4$) this occurs at high values of the temperature, i.e. when BBN begins ($T\sim 1-10$ MeV). On the other hand, for low or negative values of $\gamma$, $H^2$ becomes negative at low temperature, i.e. when BBN finishes ($T\sim 1 \,\rm keV$). Anyway, for all values of $\gamma$, there are problems when $\lambda < 0$. We could be able to study particular cases of $\gamma$ where $\lambda < 0$ but, since we are interested in a systematic study, we will concentrate only on positive values of $\lambda$.

We have modified the standard numerical code \cite{Kawano}, performing the modification (\ref{fried_mod_1}) and we have obtained the new predicted values for the several light element yields. An input that is needed for our calculation is the baryon-to-photon number ratio $\eta_B \equiv n_B/n_\gamma$. We use values consistent with the analysis of the cosmic microwave background anisotropies, where the parameter $\eta_B$ plays a role \cite{CMB, Sarkar:2002er}.

\begin{figure}[b]
  \centering
  \subfigure[]{
    \includegraphics[width=7.5cm, height=12cm, angle=-90]{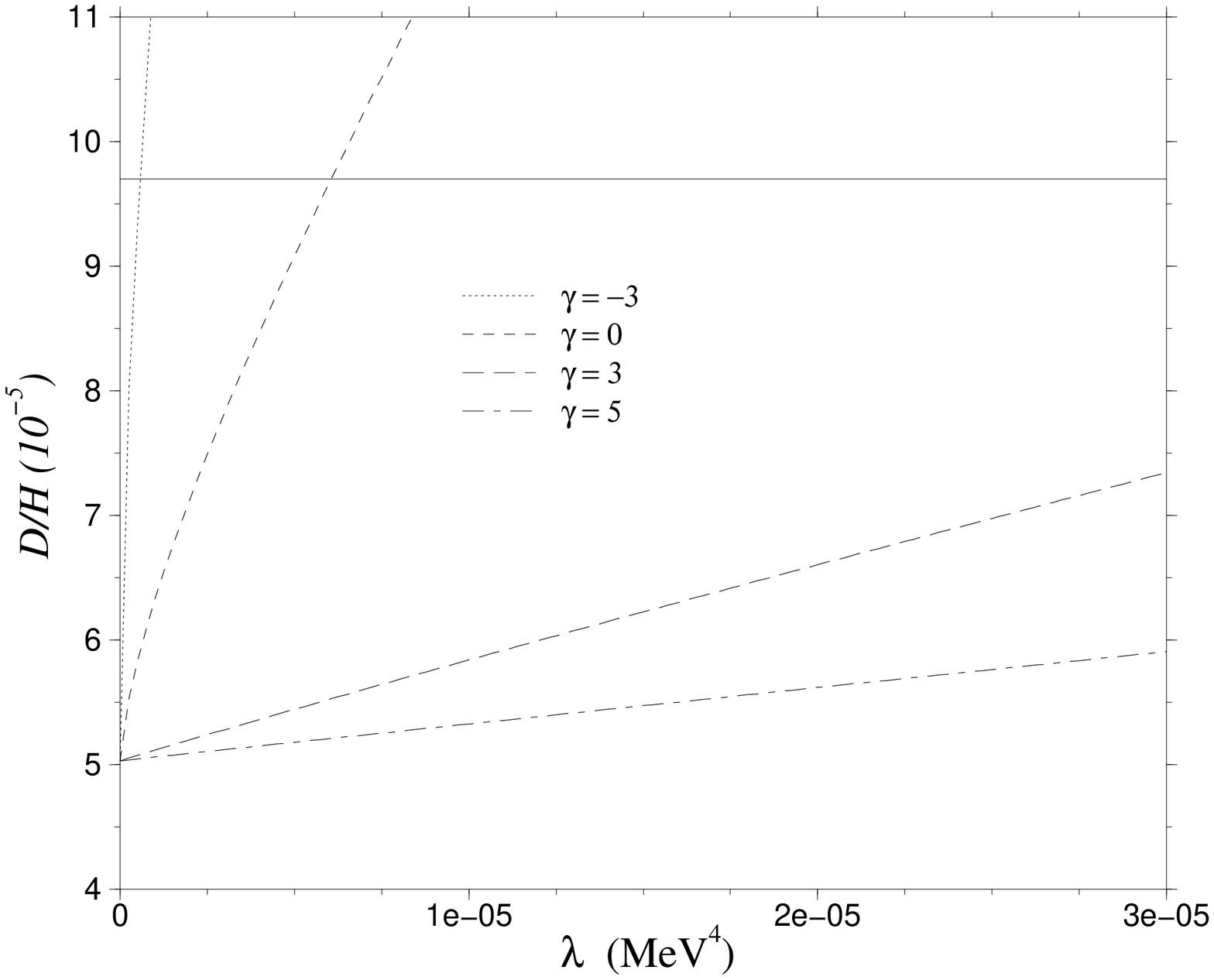}
    \label{fig1}}
  
  \subfigure[]{
    \includegraphics[width=7.5cm, height=12cm, angle=-90]{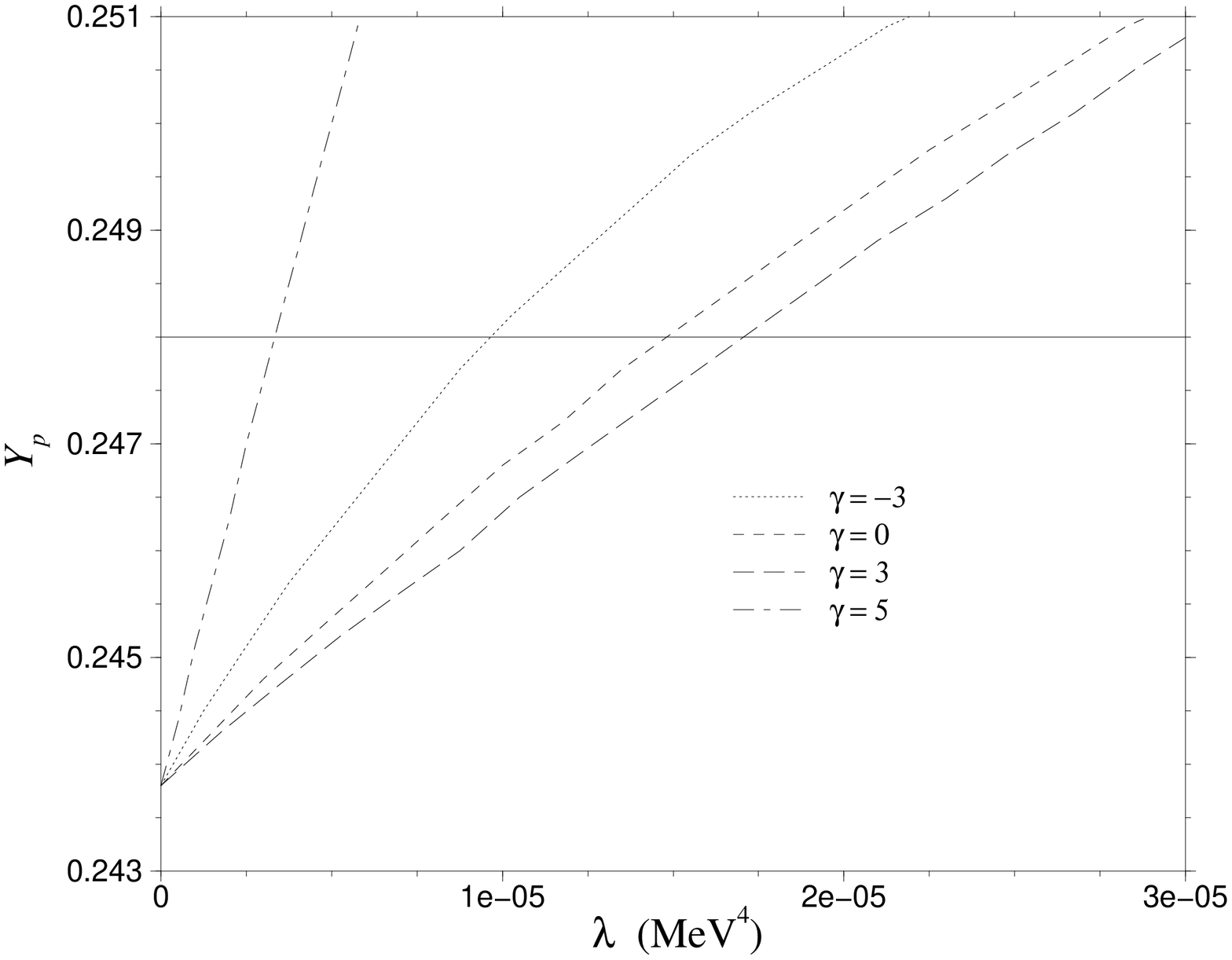}
    \label{fig2}}
  \caption{Yields of primordial deuterium, normalized to hydrogen, (a), and mass fraction of primordial helium (b), in a universe with evolution in the BBN epoch given by Eq. (\ref{fried_mod_1}), as a function of $\lambda$ and for $\gamma = -3,0,3$, and 5. We fix $\eta_B = 4\times 10^{-10}$. The horizontal line of each figure corresponds to observational data. We only show the upper limit, since the lower limit is below the displayed scale. Thus, the region above the horizontal line is excluded.}
\label{figure}
\end{figure}

In Figs. \ref{fig1} and \ref{fig2}, we can see the results for the deuterium and $^4$He abundances as a function of $\lambda$, for several values of $\gamma$. The horizontal line corresponds to the observational upper limit \cite{Sarkar:2002er}. We are presenting the results for deuterium and $^4$He because these two elements constrain more $\lambda$ and $\gamma$ than the other elements. We can see that the abundances grow with $\lambda$, in agreement with what we have reasoned before.

\begin{figure}[htb]
\begin{center}
\includegraphics[width=7.5cm, height=12cm, angle=-90]{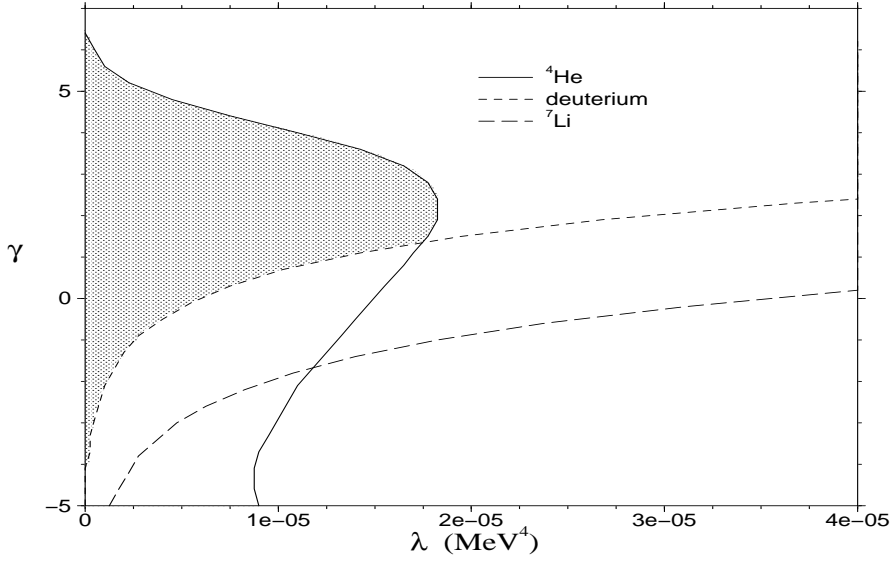}
\end{center}
\caption{\label{fig3}Constraints on the ($\gamma,\lambda$) plane coming from the requirement that the predicted abundances are consistent with the observed values, in the case $\mu_\nu = 0$. We show the constraints from $^4$He as a solid line, from D as a short-dashed line, and from $^7$Li as a long-dashed line. The intersection gives the allowed region, that we shadow. We have put $\eta_B = 4\times 10^{-10}$.} 
\end{figure}

In Fig. \ref{fig3}, the observationally allowed region for $\lambda$ and $\gamma$ is presented. Solid, dotted, and dashed lines correspond to the constraints coming from $^4$He, deuterium, and $^7$Li, respectively. The intersection of the constraints, giving the allowed region is shadowed. From this figure, we can see that the constraint coming from $^4$He is important for larger values of $\gamma$ and the one coming from deuterium is more relevant for smaller values of $\gamma$. We can understand this feature in the following way. When the nuclear reactions that produce the light elements begin, the first element formed from neutrons and protons is deuterium. Through $^3$He and tritium, all the deuterium converts into $^4$He, which is the most stable nuclide. It is when the reactions that convert deuterium into $^4$He freeze out that primordial deuterium begins to form. Therefore, primordial deuterium is formed after all the primordial $^4$He is formed. For large values of $\gamma$, the modification $\lambda \, T^\gamma$ we have introduced is important at the beginning of BBN (high temperature), i.e., during the formation of $^4$He. For small values of $\gamma$, the modification is more important at the end of BBN (low temperature), i.e., during the formation of deuterium (when all the $^4$He is already formed). This feature can also be seen in Fig. \ref{figure}. In Fig. \ref{fig1}, the maximum allowed value of $\lambda$, given by the observational constraint from deuterium, is smaller for low values of $\gamma$ and grows as $\gamma$ becomes larger. The case of $^4$He, Fig. \ref{fig2}, is similar for the low values of $\gamma$, but for $\gamma > 3$ the effect of the term $\lambda \, T^\gamma$ is more important, and the maximum allowed value of $\lambda$ is more constrained.

From Fig. \ref{fig3} we see that there are strong bounds on $\lambda$, so we conclude that departures from the FE are very much constrained. To weight the importance of the new term $\lambda\,\left(T/0.1\,\mev\right)^\gamma$ in (\ref{fried_mod_1}) we can compare it numerically with the standard contribution $\rho_o$ in (\ref{friedmann_2}). For the maximum allowed values of $\lambda$, and making the comparison at the standard BBN temperature $T=0.1 \,\mev$, we find that the new term does not exceed $10^{-1}$ times the standard term.

Carroll and Kaplinghat \cite{Carroll:2001bv} have done a similar study, motivated by the same concern than we have discussed in this section. They use the following generalization of the FE
\be
H = \left( \frac{T}{1\,\mev}\right)^\alpha \, H_1
\label{mod_carroll}
\ee
and bound the parameters $\alpha$ and $H_1$ using BBN. The form (\ref{mod_carroll}) is suitable to test radical departures from the FE as suggested by some work in brane-world models \cite{Binetruy:1999ut}. However, brane-world models developed in a series of papers \cite{Cline:1999ts} that come closer to the standard FE are, in our opinion, more promising. Thus, we think that, in this sense, our parameterization is more suitable. More support to our point of view and the form we use for the modified FE is the fact that, as it follows from our study, possibles departures from the FE have to be small due to the observational data on BBN. Even if both parameterizations are not mathematically completely equivalent, one may establish connections between the parameterization in Eq. (\ref{mod_carroll}) and ours, in Eq. (\ref{fried_mod_1}), when using suitable expansions in some limits. In this sense, our work complements the study in \cite{Carroll:2001bv}. 

Another related work has recently been done by Zahn and Zaldarriaga \cite{Zahn:2002rr}. They show how future cosmic microwave background radiation experiments will probe the FE (\ref{friedmann_1}). If we take their modification and test it in the radiation era using BBN (it would correspond to the particular case $\gamma = 4$ in our Eq. (\ref{fried_mod_1})) we reach bounds that are on the same order of magnitude that the values found in \cite{Zahn:2002rr}.

\subsection*{A Particular Case : Non-Universal Gravity}

In a recent paper \cite{Rafelski:2002gq}, Rafelski discusses a model where one would expect the fine structure constant $\alpha$ to vary in time. His work has been prompted by the recent measurements of $\alpha$ at cosmological time scales that may show this time variation effect \cite{Webb:2000mn}. He proposes that gravity couples much strongly (a factor $\sim 10^3$) to the second and third generation than it does to the first. In \cite{Rafelski:2002gq} it is claimed that the model is not excluded, since there is a lack of laboratory experimental constraints on how gravity couples to particles of the second and third generation (in the laboratory one measures the value of the gravitational strength through observations with matter of the first generation, namely electrons, protons and neutrons.)

However, the claim in \cite{Rafelski:2002gq} is partially not true. While certainly the laboratory constraints on how muons, taus, and second and third generation quarks couple to gravity are practically absent, surprisingly enough there are strong constraints on the neutrino sector. Gasperini \cite{Gasperini:zf} demonstrated that if gravity is not universally coupled to leptonic flavors, the gravitational field may contribute to neutrino oscillations. Laboratory experiments on neutrino oscillations place very strong bounds on the strength of the violation of universal gravity \cite{Gasperini:zf,Minakata:1994kt}.

BBN also offers a test of this exotic possibility, since muonic and tauonic neutrinos participate actively in the expansion of the early universe. The FE in a model of non-universal gravity reads
\be
H^2 = \frac{8\,\pi}{3}\left[G\,(\rho_\gamma + \rho_e + \rho_{\nu_e}) + G_2\,\rho_{\nu_\mu} + G_3\,\rho_{\nu_\tau} \right]
\label{fried_mod_2}
\ee
 where $G_2$ and $G_3$ are the gravitational constants for the second and third generation and $G$ is the usual Newton gravitational constant. As before, we have neglected the curvature term, which is not important in the radiation dominated epoch. 

We will now show that with some approximations we can use the general results that we have presented before in this section. First, we consider non-degenerate neutrinos (in the next section we shall drop this assumption). In this case, and if we ignore the small temperature difference between $\nu_e$, $\nu_\mu$, and $\nu_\tau$ we have $\rho_{\nu_e} = \rho_{\nu_\mu} =\rho_{\nu_\tau} \equiv \rho_\nu$, with 
\be
\rho_\nu = \frac{7}{8}\,\frac{\pi^2}{30}\,T^4
\label{rho_nu}
\ee
Then we can write (\ref{fried_mod_2}) as
\be
H^2 = \frac{8\,\pi\,G}{3}\left[\rho_o + \frac{\Delta G_2 + \Delta G_3}{G}\,\rho_\nu \right]
\label{fried_mod_3}
\ee
where $\rho_o$ is given in (\ref{rho_t}) and we have defined $\Delta G_2$ and $\Delta G_3$ as $\Delta G_i\equiv G_i - G$. Since here we are interested in getting an approximate bound on $\Delta G_2$ and $\Delta G_3$, we consider that the temperature of the neutrinos and the photons are of the same order. Thus we can see that Eq. (\ref{fried_mod_3}) is a particular case of the general modification (\ref{fried_mod_1}) with $\gamma = 4$ and 
\be
\lambda = \frac{7}{8}\,\frac{\pi^2}{30}\,\frac{\Delta G_2 + \Delta G_3}{G}\,(0.1 {\rm MeV})^4
\ee

Now, from Fig. \ref{fig3}, in the case $\gamma = 4$, we can estimate the limit $\lambda \lesssim 10^{-5}\,\rm MeV^4$, which constrains the deviation from universal gravity to values of the order
\be
\frac{\Delta G_2 + \Delta G_3}{G} \lesssim 1
\ee
But before comparing this result with what is needed in \cite{Rafelski:2002gq}, we will see in the next section that if we allow for a neutrino degeneracy, all these constraints become weaker.

\section{Testing the Friedmann equation with degenerate BBN}
\label{degenerate}

Up to this point we have taken $\mu_{\nu_i} = 0$, so we have not considered a neutrino/antineutrino asymmetry. In this section, we would like to study how the results change in the case of degenerate neutrinos, i.e., $\mu_{\nu_i} \neq 0$. To reduce the number of parameters, we put $\mu_{\nu_e} = \mu_{\nu_\mu} =\mu_{\nu_\tau} \equiv \mu_\nu$, namely that the asymmetry between neutrinos and antineutrinos is the same for the three families. Actually, it has been recently pointed out \cite{Dolgov:2002ab} that this condition is a consequence of oscillations among the different families of neutrinos. Because of the dependence of $\mu_\nu$ on the temperature, it is convenient to use the ratio $\xi_\nu \equiv \mu_\nu/T_\nu$ which is a constant quantity (again, we ignore a small temperature difference between $\nu_e$ and $\nu_\mu,\,\nu_\tau$). 

The introduction of a neutrino degeneracy strongly affects the production of $^4\rm He$. This can be easily understood because for this element, following a standard qualitative analysis \cite{Kolb:vq}, one can get an approximate analytic expression for $Y_p$, the primordial $^4\rm He$ mass fraction. It reads
\be
Y_p = \frac{2\left(n/p\right)_{Nuc}}{1 + \left(n/p\right)_{Nuc}}
\label{Y_p}
\ee
where $\left(n/p\right)_{Nuc}$ is the ratio between neutrons and protons at the moment the  $^4\rm He$ is formed. In the case of degenerate neutrinos this ratio is given by \footnote{Actually, it is the degeneracy of $\nu_e$ that matters, but we have put $\xi_{\nu_e} = \xi_{\nu_\mu} =\xi_{\nu_\tau} \equiv \xi_\nu$.}

\be
\left(n/p\right)_{Nuc} \propto \exp\left(-\frac{m_n-m_p}{T_{Nuc}}-\xi_\nu\right)
\label{n/p}
\ee
where we can see that degeneracy affects $Y_p$ exponentially. From Eqs. (\ref{Y_p}) and (\ref{n/p}) we can see that positive values of $\xi_\nu$ diminish the predicted abundance $Y_p$, an effect that goes in the opposite direction compared to the consequences  of the modification (\ref{fried_mod_1}), when $\lambda > 0$. This can be seen in Fig. \ref{fig4} where a value of $\xi_\nu = 0.06$ increases the maximum allowed value of $\lambda$ by about one order of magnitude (such value of $\xi_\nu$ is consistent with recent analysis, see \cite{Dolgov:2002ab}). As expected from our discussion above, the allowed region coming from the $^4\rm He$ restriction is significatively enlarged, while the restrictions coming from the other elements remain nearly constant. We conclude that for $\xi_\nu \neq 0$, larger values of $\lambda$ are allowed, so one has much more freedom to modificate Eq. (\ref{friedmann_2}) than in the case $\xi_\nu = 0$.

\subsection*{Constraints on Non-Universal Gravity from Degenerate BBN}

\begin{figure}[htb]
\begin{center}
\includegraphics[width=7.5cm, height=12cm, angle=-90]{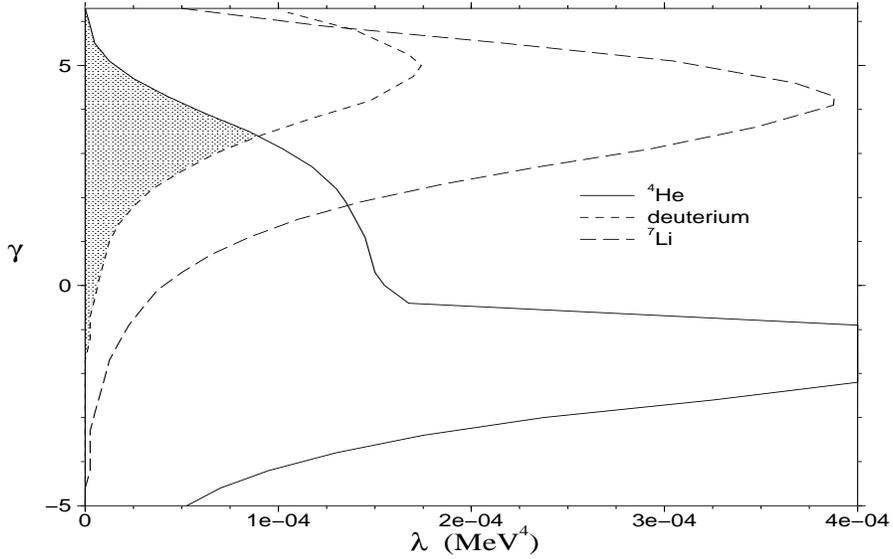}
\end{center}
\caption{\label{fig4}Same that Fig. \ref{fig3} but with $\xi_\nu = \mu_\nu/T = 0.06$. The constraints are much looser as can be seen by noticing the scale change in the $\lambda$ axis.} 
\end{figure}

In the same way as we did in the precedent section, we could treat the model of non-universal gravity as a particular case of (\ref{fried_mod_1}), and from Fig. \ref{fig4} we could limit the parameter $\frac{\Delta G_2 + \Delta G_3}{G}$ for the value $\xi_\nu = 0.06$. But here, taking profit that in this case we have fixed one parameter ($\gamma = 4$), we can make a more general analysis of how $\xi_\nu$ affects the restrictions on $\lambda$.

\begin{figure}[htb]
\begin{center}
\includegraphics[width=7.5cm, height=12cm, angle=-90]{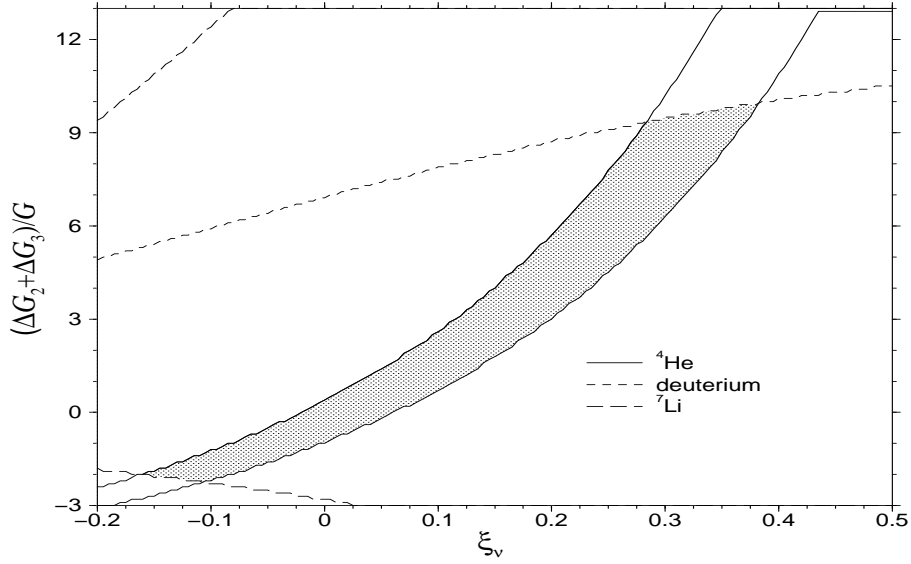}
\end{center}
\caption{\label{fig5}Region in the $\left(\frac{\Delta G_2 + \Delta G_3}{G},\xi_\nu \right)$ plane where the modified Friedmann equation leads to abundances in agreement with observation. We display the limits coming from $^4$He (solid line), D (short-dashed line), and $^7$Li (long-dashed line). The allowed region when one takes into account all constraints is shadowed. In this figure we set $\eta_B = 4\times 10^{-10}$.} 
\end{figure}

We notice that Eq. (\ref{fried_mod_3}) is still valid in the degenerate case (recall we set $\mu_{\nu_e} = \mu_{\nu_\mu} =\mu_{\nu_\tau} \equiv \mu_\nu$) but now, to get $\rho_\nu$, we have to use Eq. (\ref{rho_i}) with the temperature $T$ equal to the neutrino temperature, as it should be. We make a numerical analysis with the free parameters $\frac{\Delta G_2 + \Delta G_3}{G}$ and $\xi_\nu $. In Fig. \ref{fig5} we show the allowed region for the parameter space $\left(\frac{\Delta G_2 + \Delta G_3}{G},\xi_\nu \right)$ for the value $\eta_B = 4\times 10^{-10}$. We can see that constraints coming from the observational limits of the different elements bound these parameters into a closed region. Notice that if one only considered the restrictions coming from one element, the allowed region would be no longer closed. The reason for this is that the effects of $\frac{\Delta G_2 + \Delta G_3}{G}$ and $\xi_\nu$ on each element abundance go in opposite directions and may compensate each other. As we have noted before in the case of $^4\rm He$, positive values of $\xi_\nu$ diminish the predicted abundance $Y_p$. On the other hand, positives values of $\frac{\Delta G_2 + \Delta G_3}{G}$ increase the expansion rate of the universe, $H$, which leads to an increase of $Y_p$, compensating the change induced by $\xi_\nu$.

\begin{figure}[htb]
\begin{center}
\includegraphics[width=7.5cm, height=12cm, angle=-90]{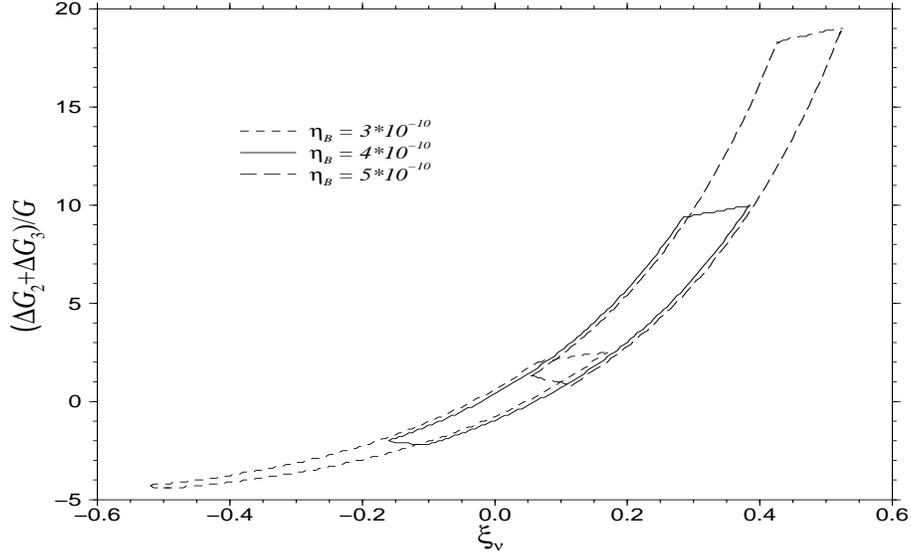}
\end{center}
\caption{\label{fig6}Allowed region in the $\left(\frac{\Delta G_2 + \Delta G_3}{G},\xi_\nu \right)$ plane. We vary $\eta_B$: we show as a solid line the case $\eta_B = 4\times 10^{-10}$ (same that in Fig. \ref{fig5}), and as a short-dashed line and as a long-dashed line the cases $\eta_B = 3\times 10^{-10}$ and $\eta_B = 5\times 10^{-10}$, respectively.} 
\end{figure}

In Fig. \ref{fig6} we show the allowed region for different values of $\eta_B$. We note  that for any $\eta_B$, there are values of $\frac{\Delta G_2 + \Delta G_3}{G}$ and $\xi_\nu$ that give theoretical predictions compatible with the primordial element observations. Rather than give accurate bounds on $\left(\frac{\Delta G_2 + \Delta G_3}{G},\xi_\nu \right)$, here we are interested in pointing out that the deviation from universal gravity is roughly limited to values $|\frac{\Delta G_2 + \Delta G_3}{G}| \lesssim 20$. At the light of this result, we conclude that the values of $G_2$ and $G_3$ that are postulated in \cite{Rafelski:2002gq}, i.e., about a factor of 1,000 bigger than the value of the Newton gravitational constant (for the first family), are excluded. 

Our bounds on non-universal gravity are much weaker than the bounds coming from oscillation experiments, which are at the level $\sim 10^{-14}$ \cite{Minakata:1994kt}. However, the bounds presented in this work may have some interest due to the following reasons. First of all, our limits and the limits from oscillations are obtained at very different times; BBN period ($z \sim 10^9$) {\it versus} today ($z = 0$). Also, there might be instances where oscillation experiments do not bound non-universal gravity. Two such cases are discussed by Gasperini \cite{Gasperini:zf}. First, if there is no mixing between neutrino weak eigenstates and the total energy eigenstates there is no gravitational induced oscillations, and thus no bounds. Second, there might be an anomalous gravitational coupling of the neutrino with respect to the charged fields but with the three $\nu$ coupling equally, namely, $\nu_e$,$\nu_\mu$, and $\nu_\tau$ all couple with strength $G'\neq G$. Then again there are no gravitationally induced oscillations. However, our limits still apply in these two cases.

\begin{acknowledgments}
We acknowledge support by the CICYT Research Project FPA2002-00648, by the EU network on Supersymmetry and the Early Universe (HPRN-CT-2000-00152), and by the \textit{Departament d'Universitats, Recerca i Societat de la Informaci{\'o}}, Project 2001SGR00188.
\end{acknowledgments}




\newpage

\end{document}